\documentstyle[epsfig]{mn}

\newif\ifAMStwofonts
\AMStwofontstrue

\def\be{\begin{equation}}
\def\ee{\end{equation}}

\def\gtsima{$\; \buildrel > \over \sim \;$}
\def\ltsima{$\; \buildrel < \over \sim \;$}
\def\prosima{$\; \buildrel \propto \over \sim \;$}
\def\gsim{\lower.5ex\hbox{\gtsima}}
\def\lsim{\lower.5ex\hbox{\ltsima}}
\def\simgt{\lower.5ex\hbox{\gtsima}}
\def\simlt{\lower.5ex\hbox{\ltsima}}
\def\simpr{\lower.5ex\hbox{\prosima}}

\title[Simulating IGM Reionization]{Simulating IGM Reionization}

\author[B. Ciardi, F. Stoehr \& S.D.M. White]{          
B. Ciardi$^1$, F. Stoehr$^{1,2}$ and S.D.M. White $^1$\\  
$^1$ Max-Planck-Institut f\"ur Astrophysik, Karl-Schwarzschild-Stra\ss e 1, 
85748 Garching, Germany\\
$^2$ Institut d'Astrophysique de Paris, 98bis Boulevard Arago, F75014 Paris,
France\\}

\date{December 2002}

\pagerange{\pageref{firstpage}--\pageref{lastpage}}
\pubyear{2002}
\begin{document}

\maketitle
\label{firstpage}

\begin{abstract}
We have studied the IGM reionization process in its full cosmological
context including structure evolution and a realistic galaxy population.
We have used a combination of high-resolution N-body simulations
(to describe the dark matter and diffuse gas component), a semi-analytic model
of galaxy formation (to track the evolution of the sources of ionizing
radiation) and the Monte Carlo
radiative transfer code {\tt CRASH} (to follow the propagation
of ionizing photons into the IGM). The process has been followed 
in the largest volume ever used for this kind of study, a field region
of the universe with a comoving length of $L \sim 20 h^{-1}$~Mpc, embedded in a much
larger cosmological simulation. 
To assess the effect of environment on the
reionization process, the same radiative transfer simulations have been
performed on a $10 h^{-1}$~Mpc comoving box, centered on a clustered region.
We find that, to account for the all ionizing radiation, objects with
total masses of $M \sim 10^9$~M$_\odot$ must be resolved. In this case, 
the simulated
stellar population produces a volume averaged ionization fraction $x_v=0.999$
by $z\sim 8$, consistent with observations without requiring any
additional sources of ionization.
We also find that environment substantially affects the reionization
process. In fact, although the simulated proto-cluster occupies a smaller volume 
and produces a higher number of ionizing photons, it gets totally
ionized later. This is because high density regions, which are more common in the 
proto-cluster, are difficult to ionize because of their high recombination rates.
\end{abstract}


\section{Introduction}

Recent discoveries of quasars at $z>5.8$ (e.g. Fan et al. 2000, 2001) are
finally allowing quantitative studies of the high-redshift
InterGalactic Medium (IGM) and its reionization history. In particular, 
the detection of a Gunn-Peterson trough (Gunn \& Peterson 1965) in
the Keck (Becker et al. 2001) and VLT (Pentericci et al. 2002) spectra of the 
Sloan Digital Sky Survey quasar SDSS 1030-0524 at $z=6.28$ is a clear 
indication that the universe is approaching the reionization epoch at 
$z\sim 6$. 

Whatever the exact value of the reionization redshift,
$z_{ion}$, it is clear that the hydrogen in the IGM is in a
highly ionized state at $z \simlt 6$. The nature of the
responsible sources is the subject of a lively debate. Several authors 
(Madau, Haardt \& Rees 1999 and references therein) have claimed that
the known populations of quasars and galaxies provide $\sim 10$ times fewer
ionizing photons than are necessary to explain the observed IGM ionization
level. Thus, additional
sources of ionizing photons are required at high redshift, the most
promising being early galaxies and quasars. While no evidence
for the existence of high-redshift quasars has yet been found, recent observations
of temperature and metal abundance in the lowest density regions of
the IGM suggest the existence of an early population of pregalactic
stellar objects, which may have contributed to the reionization and metal
enrichment of the IGM (Cowie \& Songaila 1998; Ellison et al. 2000;
Schaye et al. 2000; Madau, Ferrara \& Rees 2001). 
For this reason, most theoretical work on IGM reionization has assumed stellar sources.

The study of IGM reionization by primeval stellar sources has been
tackled by several authors, both via semi-analytic (e.g.
Haiman \& Loeb 1997; Valageas \& Silk 1999; Miralda-Escud\'{e}, Haehnelt 
\& Rees 2000; Cojazzi et al. 2000) and numerical (e.g. 
Gnedin \& Ostriker 1997; Ciardi et al. 2000; Chiu \& Ostriker 2000; Gnedin
2000; Razoumov et al. 2002) approaches. Two main ingredients are required
for a proper treatment of the reionization process: {\it i)} a 
model of galaxy formation and {\it ii)} a reliable
treatment of the radiative transfer of ionizing photons. 

The commonly accepted scenario for structure formation is of
a universe dominated by Cold Dark Matter (CDM). The evolution of
dark matter structures in such a universe is now well understood, while
the treatment of the physical processes that govern the formation and 
evolution of luminous objects (e.g. heating/cooling 
of the gas, star formation, feedback), is still quite uncertain. 
Despite many applications of hydrodynamical simulations to the structure
formation process (e.g. Cen \& Ostriker 1993, 2000; Katz, Weinberg \& 
Hernquist 1996; Katz, Hernquist \& Weinberg 1999;
Blanton et al. 1999; Pearce et al. 2001; Weinberg, Hernquist \& Katz
2002), much of our current understanding has come from semi-analytic models,
based on simplified physical assumptions. These have proved 
successful in reproducing many (but not all) observed galaxy properties
(e.g. Lacey et al. 1993; Kauffmann, White \& Guiderdoni 1993;
Heyl et al. 1995; Baugh et al. 1998; Mo, Mao \& White 1998;
Devriendt et al. 1998; Cole et al. 2000; Somerville, Primack \&
Faber 2001). Although these models do not provide as detailed
information as numerical simulations, they have the advantage
of allowing a larger dynamic range and of being fast enough to 
explore a vast range of parameters. Their main disadvantage 
is that no information on the spatial distribution of structures
is provided. For this reason a new method has recently been developed
combining N-body numerical simulations of the
dark component of matter, with semi-analytic models which `attach'
to it the galaxies and their properties (e.g. Kauffmann, Nusser \& Steinmetz 1997;
Kauffmann et al. 1999; Benson et al. 2000;
Diaferio et al. 2001; Springel et al. 2001, SWTK). The advantage of this method is
that, once the cosmology is fixed, the N-body simulation, which provides the
overall structure, has to be performed only once, while the semi-analytic
model can be run many times to explore different physical parameters describing
the galaxy formation process.

A limitation of the above method is related to the resolution of the 
N-body simulation, and turns out to be critical in simulations
aimed at the study of the reionization process. The resolution must be high
enough to follow the formation and evolution of 
the objects responsible for producing the bulk of the ionizing
radiation. At the same time, a large simulation volume
is required to have a region with ``representative''
properties and to avoid biases due to cosmic variance on small scales.
So far, simulations of reionization
have been run in fairly small volumes ($L \sim 7 h^{-1}$~Mpc
comoving; Ciardi et al. 2000; Gnedin 2000; Razoumov et al. 2002).
These fail to represent correctly the expected large-scale distribution
either of the ionizing sources or of the material to be ionized. 
The minimum mass of objects that contribute substantially to
the reionization process must be determined and then, simulations
run in the largest possible volume compatible with resolving
this mass. 
This procedure ensures the best possible treatment of the ionizing photon 
production (Ciardi 2002).

In addition to a reliable model of galaxy formation, an accurate treatment
of the propagation of ionizing photons is required to study
IGM reionization. 
The full solution of the seven dimension radiative transfer equation
is still well beyond our computational capabilities, and although in
some specific cases it is possible to reduce its dimensionality, for the
reionization process no spatial symmetry can be invoked.
Several authors (e.g. Umemura, Nakamoto \& Susa 
1999; Razoumov \& Scott 1999; Abel, Norman \& Madau 1999; Gnedin 2000;
Ciardi et al. 2001, CFMR; Gnedin \& Abel 2001; Cen 2002;
Maselli, Ferrara \& Ciardi 2003)
have recently devoted their efforts to the development of radiative
transfer codes based on a variety of approaches (e.g. ray tracing, 
Monte Carlo, local optical depth approximation). 
Given the intrinsic difficulty in solving a high dimensionality equation,
most codes aim to solve simplified problems, and cannot be
applied for more general use. Moreover,
these codes face a speed problem. Most of them are
prohibitively slow when applied in a cosmological contest.        
For this reason, approximate methods have been
developed to allow faster
calculations (e.g. Gnedin \& Abel 2001; Abel \& Wandelt 2002).

In this paper we study the IGM reionization process through
a combination of high-resolution N-body simulations (to describe
the distribution of dark matter and diffuse gas), a semi-analytic model of galaxy
formation (to track the sources of ionization) and the Monte Carlo
radiative transfer code {\tt CRASH} (to follow the propagation
of ionizing photons in the IGM; CFMR). In
Section~2 we describe the numerical simulations of structure
formation and in Section~3
the radiative transfer code. In Section~4 we present the results
of the calculation and in Section~5 and~6 our discussion and conclusions.

\section{Numerical Simulations of Structure Formation}

As discussed in the Introduction, the mass resolution of the underlying
N-body simulation plays a crucial role in modeling 
reionization. The objects that produce the bulk of the 
ionizing photons must be resolved.
While at $z>15$ the main contributors are
small mass objects ($M \sim 10^7$~M$_\odot$), at lower redshift,
the radiation from such objects is negligible compared with that
from galaxies with total (dark halo and galaxy) masses
$M>10^9$~M$_\odot$. Thus,  
our numerical simulations should be able to resolve objects
with $M \simlt 10^9$~M$_\odot$ and, at the same time, have the largest 
possible volume compatible with such mass resolution.      
Smaller mass objects remain important, however,
for the small scale clumping (Ciardi 2002; Ciardi et al. 2000).

We started out with a simulation from the VIRGO consortium ($479 h^{-1}$~Mpc
comoving on a side; Yoshida, Sheth \& Diaferio 2001) based on a $\Lambda$CDM 
cosmology with $\Omega_m = 0.3$, $\Omega_{\Lambda} = 0.7$, $h=0.7$, 
$n=1$, $\sigma_8 = 0.9$ and $\Omega_b=0.04$. 
We then selected a ``typical'' spherical region of diameter
$\sim 50 h^{-1}$~Mpc and resimulated it
four times with higher mass resolutions within the region and lower 
resolution far outside it. Particle masses in the high resolution region are 
$M_p=6.8 \times 10^{10}h^{-1}$~M$_\odot$, $4.8 \times 
10^{9}h^{-1}$~M$_\odot$, $9.5 \times 10^{8}h^{-1}$~M$_\odot$ and $1.7 \times 
10^{8}h^{-1}$~M$_\odot$ for the `M0', `M1', `M2' and `M3' simulations, 
respectively.
A discussion of the effect of mass resolution on our computations is presented 
in Section~5. All four simulations were prepared with the techniques of
SWTK and were carried out with the N-body code {\tt GADGET} (Springel,
Yoshida \& White 2001). For the highest mass resolution $\sim 7 \times 10^{7}$
particles had to be followed.
Finally, we have extracted a box of comoving side $L=20 h^{-1}$~Mpc,
which will be used to study the reionization process. A larger box ($\sim 30
h^{-1}$~Mpc) could have been extracted, but would have required a prohibitively
long radiative transfer. The choice of
a $20 h^{-1}$~Mpc box allows a reasonably fast calculation in a region of
the universe with ``representative'' properties. 
The location and mass
of the dark matter haloes in the simulation have been determined
by means of a friends-of-friends algorithm, while we identify
gravitationally bound substructures with {\tt SUBFIND} (SWTK) and build
the merging tree for haloes and subhaloes following the prescription
of SWTK. The smallest haloes have masses of
$M \sim 10^9$~M$_\odot$, consistent with the requirements described
above.

We then model the galaxy population with the semi-analytic technique of Kauffmann 
et al. (1999) in the implementation of SWTK. This procedure follows the merging
tree of the dark matter haloes extracted from the simulation, and models
galaxy formation in the haloes and subhaloes using simple recipes for gas cooling,
star formation, feedback from supernovae and galaxy-galaxy merging.
We adopt the same parameter values in these recipes as SWTK.
At the end of this process we obtain a catalogue of galaxies for each of the
51 simulation outputs, containing for each galaxy, among other quantities,
its position, stellar mass and star formation rate (SFR).
The model reproduces reasonably well the extinction and incompleteness
corrected SFR data of Somerville, Faber \& Primack (2001) up to a
redshift of 3. For higher redshifts the situation is less clear because
of the lack of observational data, but we expect the model to be accurate
to within a factor of two. The model is fully consistent with the upper
bounds from the maximal extinction correction estimates of 
Somerville, Faber \& Primack (2001).

In order to infer the emission properties of our model galaxies, we have
assumed a Salpeter IMF and a
time-dependent spectrum typical of metal-free stars (CFMR). This choice
is further discussed in Section~5. Of the emitted ionizing photons, only a
fraction $f_{esc}$ will actually be able to escape into the IGM. Throughout
the calculation we adopt $f_{esc}=5 \%$. This choice assumes that
$f_{esc}$ is independent of other physical quantities and is the same for
every galaxy. Actually, $f_{esc}$ may well vary with, e.g., redshift, mass and
structure of a galaxy, as well as with the ionizing photon production rate (e.g. Wood
\& Loeb 1999; Ricotti \& Shull 2000; Ciardi, Bianchi \& Ferrara 2002).
Given the large uncertainty in $f_{esc}$, we consider $5\%$ merely
as a reference value. This choice is further discussed in Section~5.

In order to determine the effects of environment on the reionization
process, we will also study how reionization proceeds in a proto-cluster
rather than in a field region. For this purpose we use the cluster simulations
performed by SWTK. Also in this case numerical simulations are available
at different mass resolution, `S1'-`S4'. Runs labelled with the same
numbers in the `M' and `S' series have comparable mass resolution.

\subsection{Simulation Convergence}

\begin{figure}
\psfig{figure=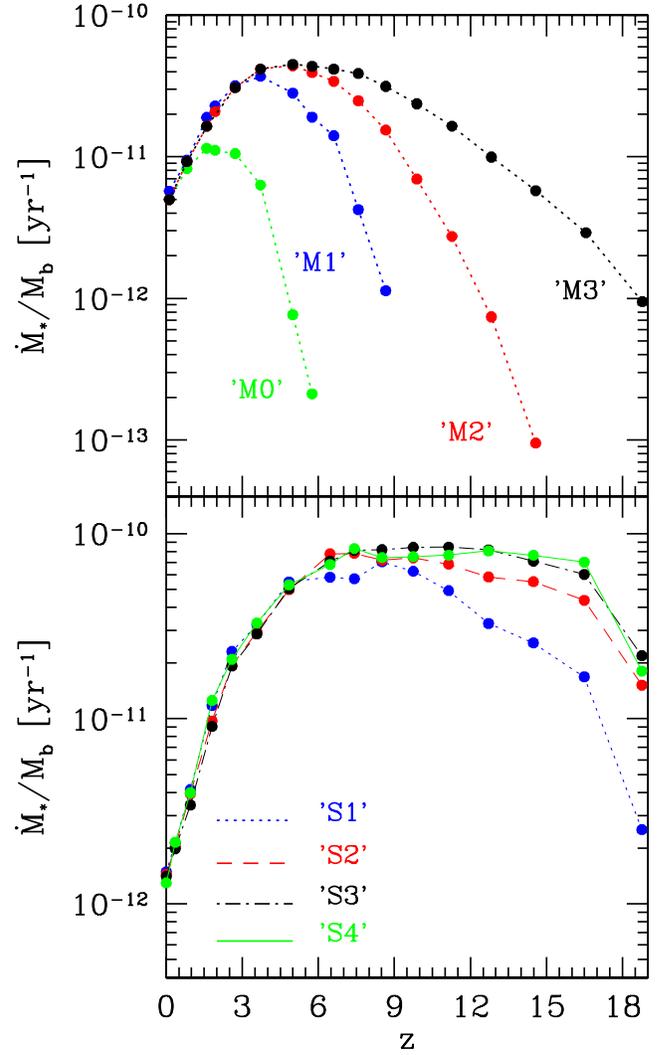,height=15cm}
\caption{\label{fig01}\footnotesize{Redshift evolution of the total star  
formation rate, $\dot{M}_\star$, normalized to the total baryonic mass, 
$M_b$, for the field region simulation (upper panel) with
four increasing mass resolutions (`M0'-`M3' from left to right).
In the lower panel the same quantity is shown for the cluster simulation.}}
\end{figure}
\begin{figure}
\psfig{figure=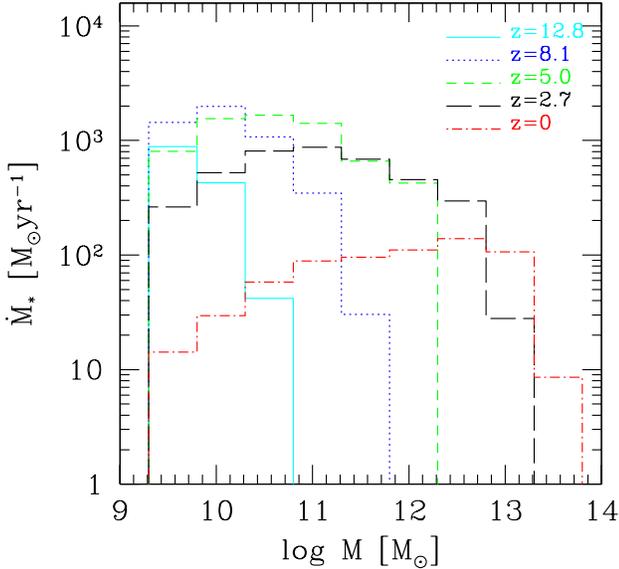,height=10cm}
\caption{\label{fig02}\footnotesize{Star formation rate, $\dot{M}_\star$,
for the `M3' simulation, as a function of the halo mass, $M$. 
The curves refer to different redshifts: $z=12.8$
(solid line), 8.1 (dotted), 5 (short dashed), 2.7 
(long dashed) and 0 (dashed-dotted).}}
\end{figure}
\begin{figure}
\psfig{figure=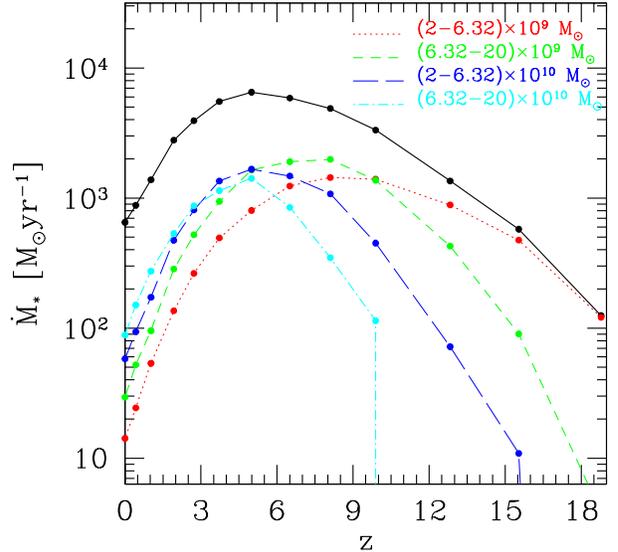,height=10cm}
\caption{\label{fig03}\footnotesize{Redshift evolution of the total star
formation rate (solid line) and the contribution from haloes with masses
in different ranges: $(2-6.32) \times 10^9$~M$_\odot$ (dotted line),
$(6.32-20) \times 10^9$~M$_\odot$ (dashed line), $(2-6.32) \times 
10^{10}$~M$_\odot$ (long dashed line) and $(6.32-20) \times 10^{10}$~M$_\odot$
(dashed-dotted line).}}
\end{figure}

We now use our series of simulations run to check for convergence.
In Fig.~\ref{fig01} the total star formation rate, $\dot{M}_\star$,
normalized to the total baryonic mass, $M_b$, is shown
as a function of redshift for the field region simulations (upper panel).
The `M2' simulation
converges to `M3' at $z_{conv}\sim 5.5$, while the `M0'
simulation converges to higher resolution results only at $z_{conv}
\sim 0$. The lower panel of Fig.~\ref{fig01} shows a similar comparison
for the cluster simulations.
In comparison to the field region,
convergence is reached at earlier times. This is due to the fact that
structure growth is accelerated in the proto-cluster region so that its
star formation becomes dominated by relatively massive galaxies much
earlier than in the field.

Thus, the `S2' and `S1' simulations converge to `S3' already at $z_{conv}
\sim 9$ and $z_{conv}\sim 8$, respectively. Comparing 
the `S3' and the `S4' simulations we conclude that the `S3',
which has a mass resolution of $M_p=2.4 \times 10^8 h^{-1}$~M$_\odot$, 
somewhat lower than `M3', already accounts for all significant star formation. 
A simulation of the field region with mass resolution comparable to `S4'
is not available. We therefore estimate the redshift after which the `M3'
run accounts for all star formation by deriving the
contribution to the total star formation from objects
with masses in different ranges. 

In Fig.~\ref{fig02} we plot $\dot{M}_\star$ as a function of the halo mass, $M$
at different redshift.  The bins are 0.5 wide, in units of log~$M$ and
they have been chosen so that the lower limit is equivalent to our mass 
resolution. Before redshift
$\sim 10$ the main contribution comes from the smallest mass objects, but
larger and larger objects become important as the redshift decreases.
This can be seen more clearly in Fig.~\ref{fig03} where the redshift evolution
of the total star formation rate is plotted (solid line) together with the
contribution from haloes with masses in different ranges. As expected, at the
higher redshifts the major contribution comes from objects with mass of few
$10^9$~M$_\odot$. At $z\sim 11$ their contribution drops below half 
of the total. 
We have plotted the contribution of haloes with mass up to $\sim 10^{11}$~M$_\odot$,
as more massive objects are important only at $z<5$, and thus are not relevant   
for our study. From this analysis we can conclude that the `M3' 
simulation accounts for most star formation at $z \simlt 11$.

\section{Radiative Transfer}

To follow the propagation of ionizing radiation produced by the
sources through the given IGM density distribution, we use the
Monte Carlo (MC) radiative transfer code {\tt CRASH} ({\it Cosmological
RAdiative transfer Scheme for Hydrodynamics}) described in CFMR.
It should be noted that in the present study a wider range of densities
is encountered, reaching values of the optical depth of a few hundreds
in the highest density regions.
For this reason, additional tests, that we do not report here, have been
run to check its reliability, finding that it reaches the same accuracy
providing a right number of photon packets is emitted (see below).
For clarity, we briefly summarize the main features of the
numerical scheme relevant to the present study.

In the application of a MC scheme to radiative transfer problems, the
radiation intensity is discretized into a representative number of
monochromatic photon packets. 
Then, all the processes involved (e.g. packet
emission and absorption) are treated statistically by randomly
sampling the appropriate distribution function and solving the time-dependent
ionization equation. Differently from CFMR, here we 
deal with more than one ionizing source. In this case, a
better solution of the discretized time-dependent ionization 
equation is obtained if the time step $\Delta t$ used in their eq.~10 is
not treated statistically. Thus, here $\Delta t$ is the time
elapsed since a photon packet has gone through the cell for which
the equation is being solved. The only inputs needed for the calculation
are the gas density field and ionization state, as well as
the source positions and emission properties. Then, {\tt CRASH}
provides the final gas ionization state, while the gas density field
remains constant throughout the radiative transfer simulation. 
The above procedure is repeated for each output of the N-body    
simulation described in the previous Section. The inputs and the {\tt CRASH}
running time, $t_{rt}$, are changed accordingly; in particular, $t_{rt}=
t_{out}$, the time between two outputs of the simulation. 
In the following, we discuss how the above inputs are derived and updated.

\begin{figure*}
\psfig{figure=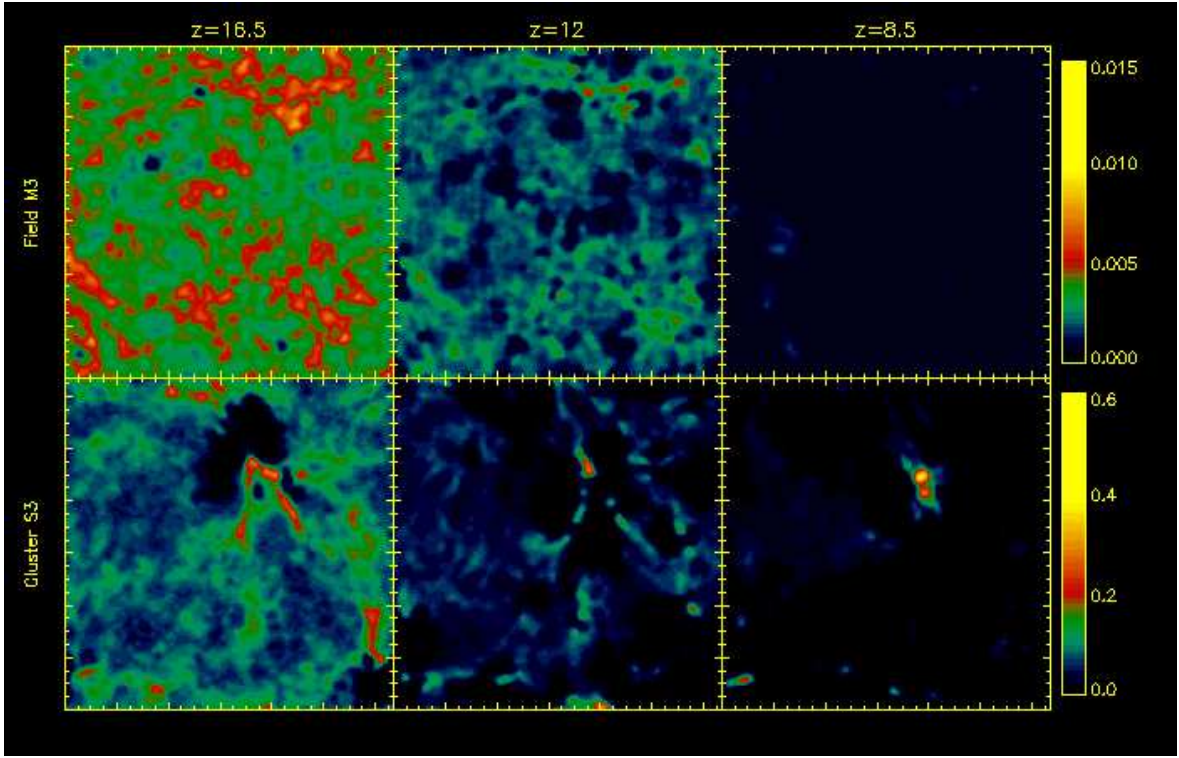,height=10cm}
\caption{\label{fig04}\footnotesize{Slices through the simulation boxes.
The six panels show the neutral hydrogen number density for the field region
(upper panels) and the proto-cluster (lower panels), at redshift, from
left to right, $z=16.5, 12$~and 8.5. The simulation box of the field
region (proto-cluster) has a comoving length of $L\sim 20 h^{-1}$~Mpc
($10 h^{-1}$~Mpc).}}
\end{figure*}

For each output of the N-body simulation, the dark matter density distribution,
$n_{dm}$, is tabulated on a mesh using a Triangular Shaped Cloud (TSC) interpolation
(Hockney \& Eastwood 1981). We discuss
the effect of neglecting density variations on sub-grid scales in Section~5.
Assuming that the gas distribution follows that    
of the dark matter, we set the gas density in each grid cell by requiring
the adimensional baryon density to be $\Omega_b=0.04$.
A number $N_c=128^3$ of cells have been used
(see Section~5 for a discussion on this choice).
As long as the sources are of stellar type, the presence of helium does not
sensibly affect the H reionization process. For this reason, we
consider a gas of pure hydrogen. Moreover, a simulation with an H/He
gas, although feasible with the updated version of the code {\tt CRASH}
(Maselli, Ferrara \& Ciardi 2003), would be extremely time consuming.
The other quantities, such as temperature, $T$,
and ionization fraction, $x$, are initialized as in CFMR 
and updated as described in the following paragraphs.

The source positions are obtained directly from the outputs of the 
N-body simulations, as described in the previous Section. In order to
reduce the computational cost of the calculation, we group all sources 
in a grid cell into a single source placed at the cell center. Mass and
luminosity conservation are assured. 
To further reduce the computational cost of the radiative transfer simulation,
each source is modeled with a number of photon packets proportional
to its luminosity, i.e. its star formation rate. 
In practice, this means that we randomly sample the source luminosity
distribution function.
For the range of densities considered here, we find that a number of photon
packets equal to ${\cal N}_p=\left [5 \times 10^7 \left (E/ 10^{60}
{\rm erg} \right ) \right ]$, where $E$ is the total ionizing energy
emitted at each output of the simulation, is needed 
to reach numerical convergence in the volume averaged ionization fraction
to within few percent. If we were rather interested, e.g., in resolving
the ionization fronts, a much higher number of photons packets would have been
required (Maselli, Ferrara \& Ciardi 2003).
Of the emitted packets, only $f_{esc}{\cal N}_p$ will actually be
able to escape from the galaxies and break into the IGM.
The packets have the same energy, ${\cal E}_p=E/{\cal N}_p$, but contain a
different number of monochromatic photons.  We determine $E$ as follows.
As ionizing photon production decreases rapidly as stars age,
the bulk of the ionizing radiation is produced by newly formed
stars. We thus assume that a mass of stars $M_\star=t_{out} \times
\sum_{i=1}^{n_s} \dot{M}_{\star,i}$, is responsible for the all emission, 
neglecting the contribution from stars formed in previous outputs. 
This is a good approximation as long as $t_{out} \simgt 10^7$~yr, the
mean lifetime of an OB star.
$\dot{M}_{\star,i}$ is the star formation rate of the {\it i}-th galaxy 
and $n_s$ is the number of galaxies at the output under consideration.
We derive $E$
integrating over $t_{out}$ the time-dependent SED of a Simple Stellar
Population of metal-free stars, with a total mass $M_\star$, distributed
according to a Salpeter IMF (see Section~2).

Given the above density distribution, source positions and emission
properties, we run the code {\tt CRASH} for the time $t_{out,k}$,
corresponding to the output, $k$, under consideration. Before running {\tt CRASH}
for the next output, $k+1$, we update the relevant physical quantities as it
follows.
Let's assume that the {\it i}-th cell has an
ionization fraction $x_i$; we thus assign to each particle in the
cell, $p_i$, the value $x_i$. We follow the same procedure for all
the cells. At the output $k+1$ the $p_i$ particles may have moved
into another cell. We thus reconstruct the new distribution of
particles, together with their
ionization fraction. Then, when we grid the box of the simulation
at the $k+1$ output, we assign to each cell an ionization fraction 
averaged over those of the particles that end up in that cell. Particles
flowing in from the external boundaries inherit the properties of the particles
in the cell. This
procedure is performed for all the quantities relevant to the
calculation, with the exception of the gas density, which is updated
from the simulations as described above. In this way, we can follow
the reionization process in an evolving universe. 
 
Finally, to mimic the presence of an unknown background, photons escaping
from a side of the box re-enter from the opposite side. At the early times,
when HII regions are still confined around their sources, few photons
exit the box. Their number increases rapidly as the regions start
to overlap. This is consistent with the expected build up of an ionizing
background (e.g. Gnedin 2000).

\section{Results}

As discussed in the previous Sections, to study the effect of the environment
on the reionization process, we have performed the
same radiative transfer simulation on a 20 $h^{-1}$ Mpc comoving box 
positioned in a field region of the universe and on a 10 $h^{-1}$
Mpc comoving box centered on a clustered region. 
It should be noticed that in the redshift range where
the radiative transfer simulations are applied, the cluster has not yet    
collapsed. The application of the {\tt CRASH} code in its first
version, that does not follow the detailed temperature evolution,
is thus appropriate, since less than 1\% of the gas is in regions with $T>>10^4$~K
(see also Mo \& White 2002). 
In the following we present the results of our simulations.
 
In Fig.~\ref{fig04} we show illustrative slices cut through the simulation boxes.
The six panels show the neutral hydrogen number density for the 
`M3' field region
(upper panels) and the `S3' proto-cluster (lower panels), at redshifts, from 
left to right, $z=16.5, 12$~and 8.5. The ionized regions   (black)
clearly grow differently in the two environments. In       
the field, high density peaks are uncommon and HII
regions easily break into the IGM; in the proto-cluster the density is
higher, so ionization is more difficult and recombination is faster. Moreover,
many photons are initially needed to ionize the high density gas 
surrounding the sources, before photons can escape into the low density
IGM. As a result, more photons are required to ionize the proto-cluster region;
although in the proto-cluster the ionizing photon production per unit mass is higher
at high redshift (see Fig.~\ref{fig01}), filaments of neutral gas are still present 
after the field region is almost completely ionized.

\begin{figure}
\psfig{figure=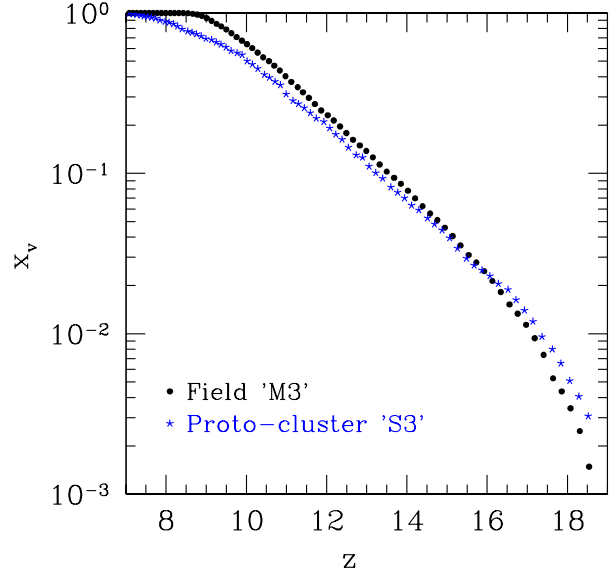,height=10cm}
\caption{\label{fig05}\footnotesize{Redshift evolution of the volume averaged
ionization fraction,
$x_v$, for the `M3' field region (filled circles) and `S3' 
proto-cluster (asterisks).}} 
\end{figure}

This is more clearly seen in Fig.~\ref{fig05}, where the volume averaged
ionization fraction, $x_v$, is shown as a function of redshift
for `M3' and `S3'. This ionization fraction is defined
as $x_v=\sum_i x_i V_i / V$, where $x_i$ and $V_i$ are the ionization
fraction and volume of the $i$-th cell respectively, and $V$ is the total 
volume of the box. The evolution proceeds differently in the two regions: 
initially the ionization fraction are comparable, but the proto-cluster 
gas ionizes more slowly, for the reasons discussed above. While the `M3' 
field region already has $x_v \sim 0.999$ at $z=8.3$, $x_v \sim 0.8$ for 
the proto-cluster at the same redshift. 

\begin{figure}
\psfig{figure=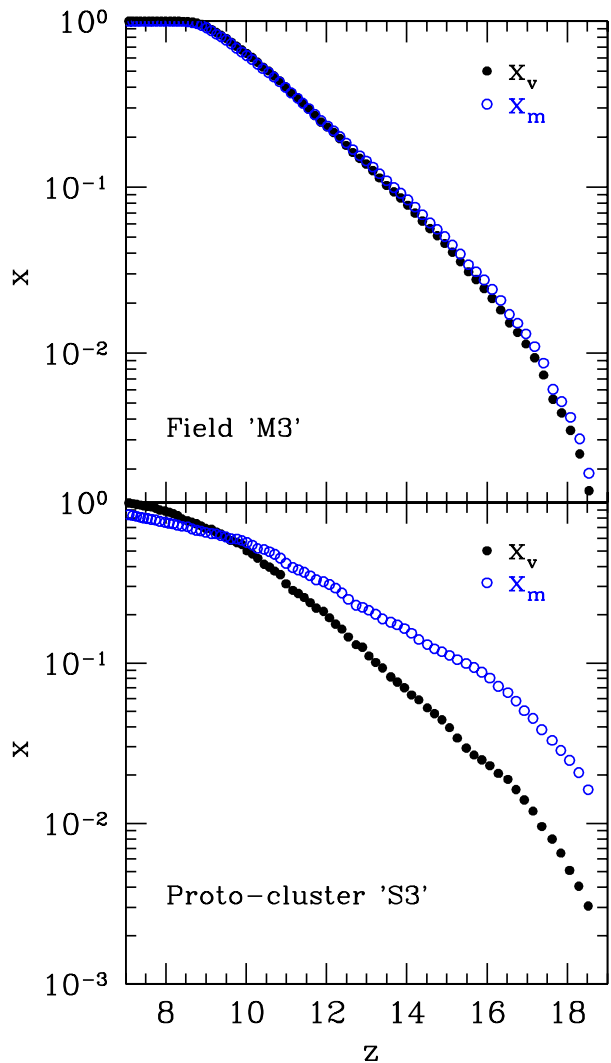,height=15cm}
\caption{\label{fig06}\footnotesize{Redshift evolution of ionization fraction
for the `M3' field region (upper panel) and the `S3' proto-cluster (lower panel).
Volume (filled circles) and mass (open circles) averaged ionization fraction
($x_v$ and $x_m$, respectively) are shown.}}
\end{figure}

A comparison between volume and mass averaged ionization fractions is useful
to understand how the reionization process proceeds. We define the latter
quantity as $x_m=\sum_i x_i M_i / M$, where $M_i$ is the mass in the 
$i$-th cell and $M$ is the total mass in the box. 
In Fig.~\ref{fig06} the redshift evolution of $x_v$ (filled circles) 
and $x_m$ (open circles) is compared both for the `M3' field region
(upper panel) and for the `S3' proto-cluster (lower panel).
The most interesting feature of these plots is that, while for the field
region $x_m$ is always comparable with $x_v$, $x_m>x_v$ initially in the
proto-cluster, i.e. at first relatively dense cells are ionized. 
This because the density immediately surrounding the sources
must be ionized, before the photons can break into the 
IGM. This indicates that the assumption adopted by some authors (e.g.
Miralda-Escud\'e, Haehnelt \& Rees 2000), that lower density gas always
gets ionized before higher density regions, fails in a cluster environment.
The trend reverses at later times, when large, low density regions get ionized
and the highest density peaks remain neutral or recombine. 
Although almost 90\% of the proto-cluster
volume is ionized by $z \sim 8$, this is true
for only $\sim 70\%$ of the mass.

\begin{figure}
\psfig{figure=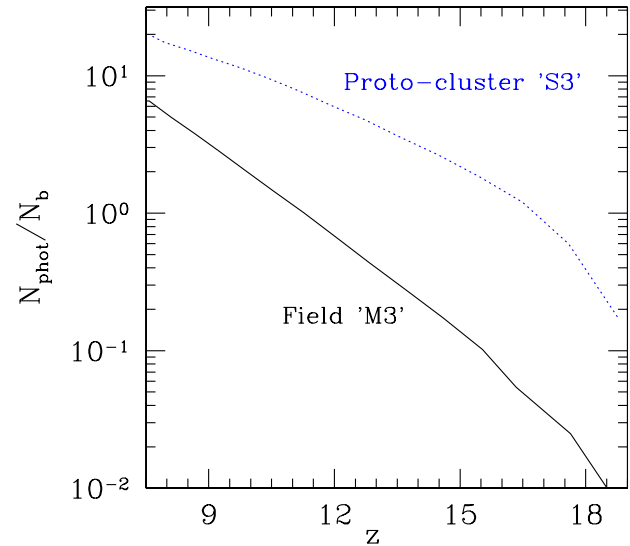,height=10cm}
\caption{\label{fig07}\footnotesize{Ratio between the cumulative number of escaping
ionizing photons, $N_{phot}$, and the number of baryons, $N_b$, as a function 
of redshift for the `M3' field region (solid line) and the
`S3' proto-cluster (dotted line).}} 
\end{figure}

Some approaches to the study of the reionization process adopt the
one-photon-per-baryon approximation. This is valid only in cases
when recombination can be safely neglected. An estimate of the photon
budget required to ionize the IGM in a more realistic case is shown
in Fig.~\ref{fig07}, where the ratio between the cumulative number of escaping 
ionizing photons, $N_{phot}$, and the number of baryons, $N_b$, is plotted as a
function of redshift for the `M3' field region (solid line) and the
`S3' proto-cluster (dotted line). High density peaks, where star
formation preferentially takes place, are more abundant in the proto-cluster,
so photon production is higher there than in the field region, at least 
up to $z\sim 11$. At $z\sim 8$, when $x_v=0.999$, 
about 5 photons per baryon have been produced in the field region. However, the
$\sim 15$ photons per baryon produced in the proto-cluster are not enough to
reionize it completely for the reasons discussed above.

\section{Discussion}

\begin{figure}
\psfig{figure=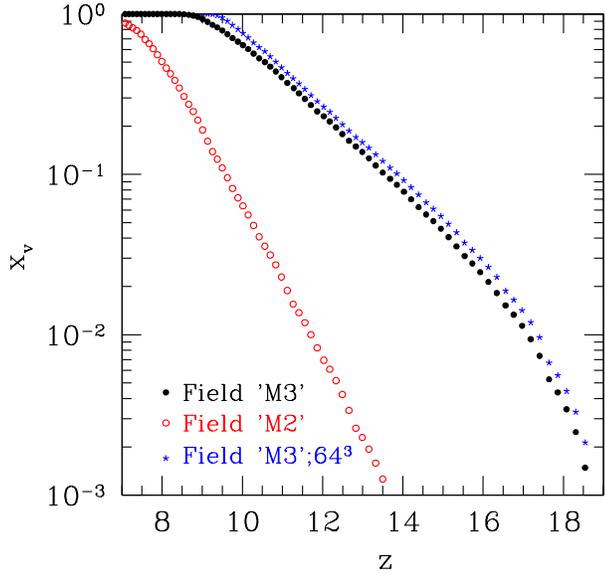,height=10cm}
\caption{\label{fig08}\footnotesize{Redshift evolution of volume 
averaged ionization fraction,
$x_v$, for the `M3' (filled circles) and the `M2' (open circles) field
region and for the `M3' field region with $N_c=64^3$ (asterisks).}}
\end{figure}

In the following we will discuss the convergence of our simulations
with respect to mass and grid resolution.
To assess the impact of mass resolution on our results, we have run
a reionization simulation on the `M2' field region. The redshift evolution of
$x_v$ is shown in Fig.~\ref{fig08} as open circles. The `M2' and `M3' 
curves converge only at $z<7$.
The difference observed at higher redshift is due to the mass resolution,
which, for the `M2' simulation, is not enough to resolve the objects that
give the main contribution to the ionizing radiation, as is evident
from the upper panel of Fig.~\ref{fig01}. 
From Figs.~\ref{fig02}-\ref{fig03} we are confident that the `M3' simulation has
the appropriate resolution to account for the bulk of ionizing photon
production, at least after $z\sim 11$. It should be noticed that 
the formation and evolution of objects with $M<10^9$~M$_\odot$ may be 
affected by feedback effects, that inhibit their star formation 
({\it e.g.} Haiman, Rees \& Loeb 1997; Omukai \& Nishi 1999; Mac Low \& 
Ferrara 1999; Barkana \& Loeb 1999; Ciardi et al. 2000; Glover \& Brand 2000;
Nishi \& Tashiro 2000; Kitayama et al. 2001; Glover \& Brand 2002; 
Benson et al. 2002). Thus, these smaller mass structures may well   
play a minor role in the reionization process.

To assess the impact of grid resolution we have rerun our radiative transfer
simulations with $N_c=64^3$. The resulting evolution of $x_v$ is 
shown in Fig.~\ref{fig08} 
by asterisks. From this figure it is evident that the grid resolution 
does not greatly affect the final results; the value of $x_v$
never differs in the two runs by  more than 
10 or $20\%$, although the case with $N_c=64^3$ 
produces slightly higher values. $N_c>128^3$ is not allowed
by the mass resolution of our N-body simulation, as sampling fluctuations
in the density field then become large.

\begin{figure}
\psfig{figure=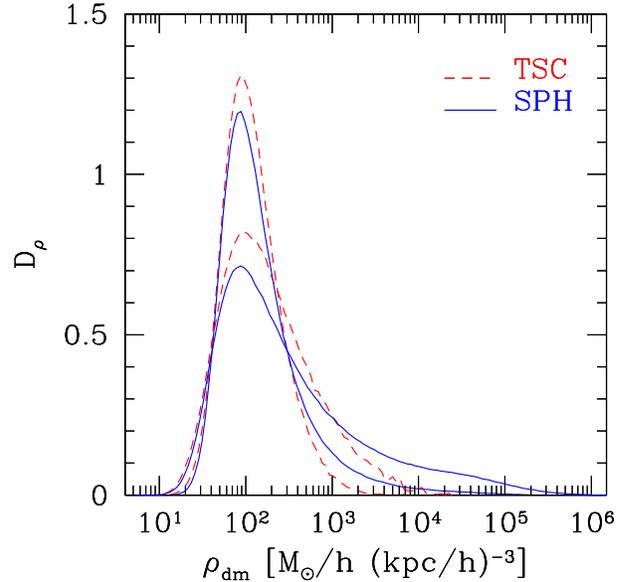,height=10cm}
\caption{\label{fig09}\footnotesize{Distribution of DM densities at the
positions of the DM particles, for a 32-particle
SPH smoothing kernel (solid line) and the TSC interpolation scheme
(dashed). The upper two
curves show the distributions for $z=10$, the lower ones for $z=6$. $D_p$
gives the relative number of particles per $\log \rho$ interval.}}
\end{figure}

As mentioned above, small mass objects, although negligible for 
ionizing photon production, are important because
small-scale clumping can enhance recombination. Simulations which do not include such
objects may underestimate the actual number of ionizing photons 
needed to reionize the IGM. To mimic clumping on unresolved 
scales, a so called clumping factor, $C$, is often introduced.
It is defined as $C=<n^2>/<n>^2$, where $n$ is the density.
The TSC technique used to derive our density distributions will lead us
to underestimate the clumping factor. The TSC interpolation has the 
advantage of producing smoothed density values at the
positions of the grid cells, as required for the radiative transfer
computation. However, this technique is not adaptive, like an SPH kernel
for example, and density variations on sub-grid scales are smoothed out,
reducing the contrast of the DM density peaks where our sources are positioned.
This results in an underestimate of the recombination rate and
thus faster reionization. In order to address the importance of this
effect we have measured densities at the particle positions using both
the TSC interpolation and a 32-particle SPH smoothing kernel. 
The result of this comparison for `M3' at redshifts $z=10$ and  $z=6$ is 
shown in Fig.~\ref{fig09}.
The overall agreement for low and intermediate densities is reasonable, 
but the smoothing of the TSC scheme is clearly visible at the high densities. 
This is especially true at $z=6$.  An estimate of the error introduced
can be quantified by the ratio of the total 
recombination rate derived with the two different schemes, $r_{SPH}/
r_{TSC}$. This ratio is equal to 6.24 (20.87) at $z=10$ (6), suggesting
that we are substantially underestimating the recombination rate in the
high density regions surrounding the sources. 
We are interested primarily in the higher redshifts, where
agreement is somewhat better. Since almost all the high density regions 
where the above corrections are important
host sources, the error in the recombination rate can be considered part
of the uncertainty on our adopted escape fraction (see below).
Moreover, the effect of clumping is alleviated by photoevaporation, which
tends to weaken the effect of gas on unresolved scales as ionization
precedes (Shapiro et al. 2003).

The number of available ionizing photons depends
on the choice of the stellar spectrum, IMF and escape fraction.
As the first stars form out of gas of primordial composition, they
are thought to be very massive, resulting in a higher ionizing
photon emission, when compared with higher metallicity stars
(e.g. Larson 1998). For example, Schneider et al. (2002) propose that
enrichment is needed to permit further fragmentation into
stars with significantly lower masses, inducing a transition from a
top-heavy to a more conventional IMF. 
The debate on the metallicity
and the IMF of the first stars is still very lively and many authors
argue that metal-free stars are distributed according to a 
top-heavy IMF. 
On the other hand, Christlieb et al. (2002) recently discovered a low mass
star with metallicity only $3 \times 10^{-6}$ that of the Sun.
To be conservative, in this paper we have assumed metal-free stars with 
a Salpeter IMF.

Even more uncertain is the value of the escape fraction. Values derived
in previous work
vary from 0 to 60\% (see e.g. Wood \& Loeb 1999; Ricotti \& Shull 2000; Ciardi,
Bianchi \& Ferrara 2002 for theoretical work and e.g. Leitherer et al. 1995;
Hurwitz et al. 1997; Bland-Hawthorn \& Maloney 1999; Steidel, Pettini \&
Adelberger 2001 for observations), according to the galaxy mass, redshift and
density distribution. 
Given these uncertainties and the impossibility of accurately modeling this
parameter for every galaxy (see e.g. Ciardi, Bianchi \& Ferrara 2002 for a 
discussion), we regard $f_{esc}=5\%$ as a constant, reference value. It should
be noticed that, as the resolution of the radiative transfer calculation
is equivalent to the dimension of a single grid cell (which is always 
larger than the dimension of the sources), $f_{esc}$ is actually 
an ``effective'' escape fraction, i.e. it includes both the contribution
to the absorption of the interstellar medium in the emitting object
(``real'' escape fraction) and the one of the IGM in the same
cell as the object.

Although it is not possible to disentangle different choices for the source 
emission properties from clumping
on small scales, their expected global effect
can be estimated introducing the parameter ${\cal A}=C_n^{-1} (f_{esc}/0.05)
(N_\gamma/1\times 10^{61} {\rm M}_\odot^{-1})$, where $C_n$ is the clumping
factor normalized to the value of our simulations (e.g. $C_n=2$ means
that the clumping factor is twice that of the simulations) and
$N_\gamma$ is the total number of ionizing photons per unit solar mass of
formed stars produced by the adopted spectrum and IMF. $N_\gamma$ is 
normalized to a Salpeter IMF and a spectrum typical of metal-free stars
(CFMR). From a ``back of the envelope'' calculation, we conclude
that if ${\cal A}=10$, the $\sim 5$ photons per baryon required to
reionize the IGM are already available at $z \sim 13$, while if 
${\cal A}=0.1$ the above condition is never reached. 
As the most recent observations of high redshift quasars
(Becker et al. 2001; Djorgovski et al. 2001; Pentericci et al. 2002)
suggest that the universe was already ionized at $z \sim 6.3$, the lowest
value of ${\cal A}$ consistent with this constraint is $\sim 0.7$. A lower value
of ${\cal A}$ would require additional sources of ionizing photons to
produce the observed ionization level. In a
forthcoming paper we will discuss the possible contribution to ionizing
radiation from a primordial population of quasars (see also Wyithe \& 
Loeb 2002). In a recent paper
Ricotti (2002) has suggested that a possible contribution to ionizing
photons could also come from globular clusters.
Given our conservative assumptions on the value of the escape fraction and
the IMF, a higher value of the clumping factor would also be compatible
with a reionization driven by primordial stellar sources.

Recent results from the {\it WMAP} satellite require a mean optical depth
to Thomson scattering $\tau_e \sim 0.17$, suggesting that reionization
must have begun at relatively high redshift (e.g. Kogut et al. 2003; Spergel
et al. 2003). Reconciling an early reionization with observations of the
Gunn-Peterson effect in $z>6$ quasars, which imply a volume-averaged neutral
fraction above $10^{-3}$ at $z \sim 6$, may be challenging.
In Ciardi, Ferrara \& White (2003) we run additional
simulations and discuss our model in view of these new data.

\section{Conclusions}

We have studied the IGM reionization process in a full cosmological
context with structure evolution and a realistic galaxy population.
We have used a combination of high-resolution N-body simulations
(to describe the dark component of matter), a semi-analytic model
of galaxy formation (to track the gas evolution) and the Monte Carlo
radiative transfer code {\tt CRASH} (to follow the propagation
of ionizing photons into the IGM; CFMR). The process has been followed
in the largest volume ever used for this kind of study, with a comoving
length of $L \sim 20 h^{-1}$~Mpc, a mass resolution of $\sim 10^8
$~M$_\odot$ and tree from periodic boundary conditions. This allows us to 
resolve the primary objects responsible for the production of ionizing 
radiation and, at the same time, to simulate
a region of the universe with ``mean'' properties. To study the effect
of environment on the reionization process, we have performed the same
radiative transfer simulation on a $10 h^{-1}$~Mpc comoving box, centered
on a proto-cluster region. The main results discussed in this
paper can be summarized as follows.

(1) Our field region of the universe gets reionized by the simulated
galaxy population at a redshift $z_{ion} \sim 8$, while the proto-cluster,
although it produces a higher number
of ionizing photons, gets ionized later. This is due to the fact
that high density gas, which is more common in the proto-cluster, is  more 
difficult to ionize and recombines much faster.

(2) While for the field region $\sim 5$ photons per baryon
are enough to produce $x_v=0.999$ at $z\sim 8$, the $\sim 15$ photons per
baryon produced by this time in the proto-cluster are not enough to 
reionize it.

(3) For the same reason, while for the field region the mass and volume 
averaged ionization fractions, $x_m$ and $x_v$ respectively, are always 
comparable, $x_m$ substantially exceeds $x_v$ for the proto-cluster, at 
high redshifts. This is because
the high density regions surrounding the sources have to be 
ionized first, before the photons can break out into the low density IGM.

(4) The primordial stellar sources considered in this study give a value
of the reionization epoch consistent with observation, without
invoking the presence of additional sources of ionization.  Our conservative
assumptions on the value of the escape fraction and the IMF can
accommodate the additional clumping which remains unresolved in our scheme. 

(5) The mass resolution of the simulations can deeply affect the final
results. Objects with total masses of $M \sim 10^9$~M$_\odot$ or less must be 
resolved to account for the bulk of the star formation.

\section*{Acknowledgments}

We would like to thank V. Springel for providing us with the cluster 
simulations and an anonymous referee for his/her useful comments. 
B.C. is grateful to F.S. for his patience with computer
related questions and A. Ferrara for useful comments. This 
work has been partially supported by the Research and Training Network
``The Physics of the Intergalactic Medium'' set up by the European
Community under the contract HPRN-CT-2000-00126.

\label{lastpage}

\newpage


\begin{thebibliography}{99}

\bibitem{} Abel, T., Norman, M. L. \& Madau, P. 1999, ApJ, 523, 66
\bibitem{} Abel, T. \& Wandelt, B. D. 2002, MNRAS, 330, L53
\bibitem{} Barkana, R. \& Loeb, A. 1999, ApJ, 523, 54
\bibitem{} Baugh, C. M., Cole, S., Frenk, C. S., Lacey, C. G., 1998, ApJ, 498, 504
\bibitem{} Becker, R. H. et al. 2001, AJ, 122, 2850
\bibitem{} Benson, A. J., Baugh, C. M., Cole, S., Frenk, C. S. \& Lacey, C. G. 2000,
           MNRAS, 316, 107 
\bibitem{} Benson, A. J., Lacey, C. G., Baugh, C. M., Cole, S. \& Frenk, C. S. 2002,
           MNRAS, 333, 156
\bibitem{} Bland-Hawthorn, J. \& Maloney, P. R. 1999, ApJ, 510, 33
\bibitem{} Blanton, M., Cen, R., Ostriker, J. P. \& Strauss, M. 1999, ApJ, 522, 590
\bibitem{} Cen, R. 2002, ApJS, 141, 211
\bibitem{} Cen, R. \& Ostriker, J. P. 1993, ApJ, 417, 415
\bibitem{} Cen, R. \& Ostriker, J. P. 2000, ApJ, 538, 83
\bibitem{} Chiu, W. A. \& Ostriker, J. P. 2000, ApJ, 534, 507
\bibitem{} Ciardi, B. 2002, in ''The Evolution of Galaxies. II. Basic Building Blocks''
           proceedings, Ile de la Reunion, France, October 16-21 2001; eds. M. Sauvage, 
           G. Stazinska \& D. Schaerer; Kluwer Academic Publishers; pp. 515-518
\bibitem{} Ciardi, B., Bianchi, S. \& Ferrara, A. 2002, MNRAS, 331, 463
\bibitem{} Ciardi, B., Ferrara, A., Governato, F. \& Jenkins, A. 2000, MNRAS, 314, 611
\bibitem{} Ciardi, B., Ferrara, A., Marri, S. \& Raimondo, G. 2001, MNRAS, 324, 381 (CFMR)
\bibitem{} Ciardi, B., Ferrara, A. \& White, S. D. M. 2003, astro-ph/0302451
\bibitem{} Cojazzi, P., Bressan, P., Lucchin, F., Pantano, O. \& Chavez, M. 2000, MNRAS,
           315, 51
\bibitem{} Cole, S., Lacey, C. G., Baugh, C. M. \& Frenk, C. S. 2000, MNRAS, 319, 168
\bibitem{} Cowie, L. L., \& Songaila, A. 1998, Nature, 344, 44
\bibitem{} Christlieb, N. et al. 2002, Nature, 419, 904
\bibitem{} Devriendt, J. E. S., Sethi, S. K., Guiderdoni, B. \& Nath, B. B. 1998,
           MNRAS, 298, 708
\bibitem{} Diaferio, A. et al. 2001, MNRAS, 323, 999
\bibitem{} Djorgovski, S. G., Castro, S., Stern, D. \& Mahabal, A. A. 2001, ApJ, 560, L5
\bibitem{} Ellison, S. L., Songaila, A., Schaye, J. \& Pettini, M. 2000, AJ, 120, 1175
\bibitem{} Fan, X. et al. 2000, AJ, 120, 1167
\bibitem{} Fan, X. et al. 2001, AJ, 122, 2833
\bibitem{} Glover, S. C. O. \& Brand, P. W. J. L. 2000, MNRAS, 321, 385
\bibitem{} Glover, S. C. O. \& Brand, P. W. J. L. 2002, astro-ph/0205308
\bibitem{} Gnedin, N. Y. 2000, ApJ, 535, 530
\bibitem{} Gnedin, N. Y. \& Abel, T. 2001, NewA, 6, 437
\bibitem{} Gnedin, N. Y. \& Ostriker, J. P. 1997, ApJ, 486, 581
\bibitem{} Gunn, J.E. \& Peterson, B.A. 1965, ApJ, 142, 1633 
\bibitem{} Haiman, Z. \& Loeb, A. 1997, ApJ, 483, 21
\bibitem{} Haiman, Z., Rees, M. J. \& Loeb, A. 1997, ApJ, 476, 458
\bibitem{} Heyl, J. S., Cole, S., Frenk, C. S. \& Navarro, J. F. 1995, MNRAS, 274, 755
\bibitem{} Hockney, R. W. \& Eastwood, J. W. 1981, in ''Computer Simulations
           Using Particles''; New York: McGrow-Hill
\bibitem{} Hurwitz, M., Jelinsky, P. \& Dixon, W. V. 1997, ApJ, 481, L31
\bibitem{} Katz, N., Hernquist, L.\& Weinberg, D. H. 1999, ApJ, 523, 463
\bibitem{} Katz, N., Weinberg, D. H.\& Hernquist, L. 1996, ApJS, 105, 19
\bibitem{} Kauffmann, G., Colberg, J. M., Diaferio, A. \& White, S. D. M. 1999,
           MNRAS, 303, 188
\bibitem{} Kauffmann, G., Nusser, A. \& Steinmetz, M. 1997, ApJ, 286, 795
\bibitem{} Kauffmann, G., White, S. D. M. \& Guiderdoni, B. 1993, MNRAS, 264, 201
\bibitem{} Kitayama, T., Susa, H., Umemura, M. \& Ikeuchi, S. 2001, MNRAS, 326, 1353
\bibitem{} Kogut, A. et al. 2003, astro-ph/0302213
\bibitem{} Lacey, C., Guiderdoni, B., Rocca-Volmerange, B. \& Silk, J. 1993, 
           ApJ, 402, L15
\bibitem{} Larson, R. B. 1998, MNRAS, 301, 569
\bibitem{} Leitherer, C. et al. 1999, ApJS, 123, 3
\bibitem{} Mac Low, M.-M. \& Ferrara, A. 1999, ApJ, 513, 142
\bibitem{} Madau, P., Ferrara, A. \& Rees, M. J. 2001, ApJ, 555, 92
\bibitem{} Madau, P., Haardt, F. \& Rees, M. J. 1999, ApJ, 514, 648
\bibitem{} Maselli, A., Ferrara, A. \& Ciardi, B. 2003, in prep.
\bibitem{} Miralda-Escud\'e, J., Haehnelt, M. \& Rees, M. R. 2000, ApJ, 530, 1
\bibitem{} Mo, H. J., Mao, S. \& White, S. D. M. 1998, MNRAS, 295, 319
\bibitem{} Mo, H. J. \& White, S. D. M. 2002, MNRAS, 336, 112
\bibitem{} Nishi, R. \& Tashiro, M. 2000, ApJ, 537, 50
\bibitem{} Omukai, K. \& Nishi, R. 1999, ApJ, 518, 64
\bibitem{} Pearce, F. R. et al. 2001, MNRAS, 326, 649
\bibitem{} Pentericci et al. 2002, ApJ, 123, 2151
\bibitem{} Razoumov, A. O., Norman, M. L., Abel, T. \& Scott, D. 2002, ApJ, 572, 695
\bibitem{} Razoumov, A. \& Scott, D. 1999, MNRAS, 309, 287
\bibitem{} Ricotti, M. 2002, astro-ph/0208352
\bibitem{} Ricotti, M. \& Shull, J. M. 2000, ApJ, 542, 548
\bibitem{} Schaye, J., Rauch, M., Sargent, W. L. W. \& Kim, T. 2000, ApJ, 541, L1
\bibitem{} Schneider, R., Ferrara, A., Natarajan, P. \& Omukai, K. 2002,
           ApJ, 571, 30
\bibitem{} Shapiro, P. R., Iliev, I. T., Raga, A. C. \& Martel, H. 2003, astro-ph/0302339
\bibitem{} Somerville, R. S., Primack, J. R. \& Faber, S. M. 2001, MNRAS, 320, 504
\bibitem{} Spergel, D.N. et al. 2003, astro-ph/0302207
\bibitem{} Springel, V., White, S. D. M., Tormen, G. \& Kauffmann, G. 2001, MNRAS,
           328, 726 (SWTK)
\bibitem{} Springel, V., Yoshida, N. \& White, S. D. M. 2001, NewA, 6, 79
\bibitem{} Steidel, C. C., Pettini, M. \& Adelberger, K. L. 2001, ApJ, 546, 665
\bibitem{} Umemura, M., Nakamoto, T. \& Susa, H. 1999, in Numerical Astrophysics,
           eds.Miyama et al. (Kluwer: Dordrecht), p.43
\bibitem{} Valageas, P. \& Silk, J. 1999, A\&A, 347, 1
\bibitem{} Weinberg, D. H., Hernquist, L.\& Katz, N. 2002, ApJ, 571, 15
\bibitem{} Wyithe, J. S. B. \& Loeb, A. 2002, astro-ph/0209056
\bibitem{} Wood, K. \& Loeb, A. 1999, AAS, 195, 1309
\bibitem{} Yoshida, N., Sheth, R. K. \& Diaferio, A. 2001, MNRAS, 328, 669

\end{thebibliography}
\end{document}